\documentclass{article}

\usepackage{arxiv}

\usepackage[utf8]{inputenc} 
\usepackage[T1]{fontenc}    
\usepackage{hyperref}       
\usepackage{url}            
\usepackage{booktabs}       
\usepackage{amsfonts}       
\usepackage{nicefrac}       
\usepackage{microtype}      
\usepackage{lipsum}

\usepackage[ngerman,english]{babel}             
\usepackage[fixlanguage]{babelbib}              
\usepackage{amsmath}							
\usepackage[version=3]{mhchem}                  
\usepackage{graphicx}
\usepackage{xcolor}								
\usepackage{algorithm}							
\usepackage[noend]{algpseudocode}

\title{Deflagration-to-detonation transition in an unconfined space: II. \\ Expanding hydrogen-oxygen flames}

\author{
  Andrey Koksharov \\
  School of Mathematical Sciences\\
  Sackler Faculty of Exact Sciences \\
  Tel Aviv University\\
  Tel Aviv 69978 \\
  \texttt{koksharov@gmx.net} \\
   \And
  Leonid Kagan \\
  School of Mathematical Sciences\\
  Sackler Faculty of Exact Sciences \\
  Tel Aviv University\\
  Tel Aviv 69978 \\
  \texttt{kaganleo@tauex.tau.ac.il} \\
   \And
  Gregory Sivashinsky
	\\
  School of Mathematical Sciences\\
  Sackler Faculty of Exact Sciences \\
  Tel Aviv University\\
  Tel Aviv 69978 \\
  \texttt{grishas@tauex.tau.ac.il} \\
}

\begin{document}
\maketitle

\begin{abstract}
The problem of deflagration-to-detonation transition in an unconfined environment is revisited.  With a freely expanding self-accelerating hydrogen-oxygen flame as an example, it is shown that deflagration-to-detonation transition is indeed possible provided the flame is large enough.  The transition occurs prior to merging of the flame with the flame-supported precursor shock.  The pre-transition flame does not reach the threshold of CJ-deflagration.  Numerical simulations employed are based on the recently developed pseudo-spectral method with time and space adaptation.  
\end{abstract}


\section{Introduction}
In part 1 of this article \cite{koksharov18deflagration} it is shown that a freely expanding self-accelerating spherical flame governed by a one-step Arrhenius kinetics coupled with Darrieus-Landau wrinkling undergoes an abrupt transition to detonation (DDT) when the flame radius $R_f$ reaches a critical value, $R_{ddt}$.  In the present paper to assess $R_{ddt}$  for a realistic multistep chemistry and multicomponent transport we explore the case of a stoichiometric hydrogen-oxygen system,  \ce{2H2 + O2}, initially at $T_0 = 300$\,K and $P_0 = 1$\,bar.

The results obtained substantiate qualitative predictions of \cite{koksharov18deflagration}, yielding
$R_{ddt} = 10$\,m noticeably behind the precursor shock (s) for which 
$R_{s,ddt} = 11.5$\,m.

\section{Mathematical model}

In spherical symmetry, the set of governing equations for state variables velocity in radial direction $u$, temperature $T$ and concentrations of reactants $c_i$ reads as

\textbf{Momentum conservation}
\begin{equation}
	\frac{\partial}{\partial t} \left(r^2 \, \rho u \right)
	+ \frac{\partial}{\partial r}\bigg( r^2 \left( \rho u^2 + \tau_{r r} \right) \bigg)
	\,=\,
	 -r^2 \frac{\partial P}{\partial r}
	+ r \left( \tau_{\theta \theta} + \tau_{\varphi \varphi}\right),
	\label{eq:cons_impuls_1d}
\end{equation}

\textbf{Energy conservation}
\begin{equation}
	\frac{\partial}{\partial t}\left(r^2 \, \rho e\right)
	+ \frac{\partial}{\partial r} \Bigg( r^2 \bigg( u \left( \rho e + P + \tau_{r r} \right) + \sum_{i=1}^{n_s} {h}_i {j}^d_{i} - \lambda \frac{\partial T}{\partial r} \bigg) \Bigg)
	= 0,
\label{eq:cons_energy_1d}
\end{equation}

\textbf{Species  conservation}
\begin{equation}
	\frac{\partial}{\partial t}\left(r^2 \, c_i \right)
	+ \frac{\partial}{\partial r} \bigg( r^2 \left( u c_i + {j}^d_{i} \right) \bigg)
	= r^2 \, \Sigma^2 \dot{\omega}_i, ~~~ i=1,...,n_s,
\label{eq:cons_species_1d}
\end{equation}

Here $n_s$ -- number of species,	 $\dot{\omega}_i$ -- chemical source of species, ${j}^d_{i}$ -- molar diffusion flux,	 ${h}_i$ -- molar enthalpy, and $\Sigma$ is the degree of folding \cite{deshaies1989flame} -- the ratio of the total area of the wrinkled front to the area associated with its average radius $R_f$.
The reason for $\Sigma^2$ in Eq.~\ref{eq:cons_species_1d} is explained as follows.  According to the classical Zel'dovich--Frank-Kamenetskii theory \cite{zeldovich1985}, for a low Mach number planar flame its propagation velocity relative to the gas is proportional to the square root of the reaction rate.  On the other hand, the effective velocity of the wrinkled flame is proportional to its degree of folding $\Sigma$. Hence, the effective reaction rate of the wrinkled flame should be proportional to $\Sigma^2$.  Indeed, simulations of the $\Sigma^2$ -- based model corroborate this assessment for moderately high $\Sigma$-s (see \cite{kagan2017parametric}, \cite{koksharov18deflagration} and Fig.~5 below). For general Mach numbers $\Sigma^2$-model is clearly an extrapolation, hopefully providing a reasonably good description of the physics involved. 

The following identities are used to close the above equation system.
\begin{itemize}
	\item 
	Components of shear stress tensors:
	\begin{equation}
	\left\{
		\begin{array}{rl}
			\tau_{r r} &= -\dfrac{4}{3} \mu \left( \dfrac{\partial u}{\partial r} - \dfrac{u}{r} \right), \\
			\tau_{\theta\theta} &= \dfrac{2}{3} \mu \left( \dfrac{\partial u}{\partial r} - \dfrac{u}{r}\right), \\
			\tau_{\varphi\varphi} &= \dfrac{2}{3} \mu \left( \dfrac{\partial u}{\partial r} - \dfrac{u}{r}\right). \\
		\end{array}
	\right.
	\end{equation}
	\item
	Density
	\begin{equation}
		\rho = \sum_{i=1}^{n_s} W_i c_i
		\label{eq:dens}
	\end{equation}
	\item
	Total specific energy:
	\begin{equation}
		\rho e = \sum_{i=1}^{n_s} c_i \left( h_i + \frac{1}{2} u^2 W_i \right) - P,
		\label{eq:def_energy}
	\end{equation}
	where ${h}_i$ is the molar enthalpy and $W_i$ is the molar mass of species $i$.
	\item
	Equation of state for ideal gas:
	\begin{equation}
		P = R^0 T \sum_{i=1}^{n_s} c_i,
	\end{equation}
	where $R^0$ is the universal gas constant.	
\end{itemize}

Here $W_i$ denotes the molar mass of species $i$.

Coefficients of dynamic viscosity $\mu$, heat conductivity $\lambda$ and molar diffusion flux $\mathrm{j}^d_{i}$ are computed according to the mixture averaged formulation \cite{bird07}:
\begin{itemize}
\item 
	Dynamic viscosity of gas mixture
	\begin{equation}
		\mu = \sum\limits_i \frac{X_i \mu_i}{\sum\limits_j X_j \Phi_{ij}},
		\label{eq:viscosity_mix}
	\end{equation}
\item 
	Heat conductivity 
	\begin{equation}
		\lambda = \sum\limits_i \frac{X_i \lambda_i}{\sum\limits_j X_j \Phi_{ij}}.
		\label{eq:heatcond_mix}
	\end{equation}
\item
	Molar diffusion flux
	\begin{equation}
		{\mathrm{j}^d}_{i} = - \sum_j D^c_{ij} \frac{\partial c_j}{\partial r}
	\end{equation}
\item
	Diffusion coefficient for concentrations, with correction for mass conservation $\sum_i W_i {j}^{d}_{i} = 0$ (see e.g. \cite{kee03})
	\begin{equation}
		D^c_{ij} = X_i \left[ \left(\delta_{ij} - X_i - Y_i \right) 	D^*_i
			+ \sum\limits_{k}Y_k D^*_k X_k \right],
	\end{equation}	
	
\end{itemize}

Here the dimensionless factor, used for calculation of viscosity and heat conductivity coefficient is
\begin{equation}
\Phi_{ij} = \frac{1}{\sqrt{8}}\left( 1 + \frac{W_i}{W_j} \right)^{-1/2}
 \left[ 1 + \left( \frac{\mu_i}{\mu_j} \right)^{1/2} \left( \frac{W_j}{W_i} \right)^{1/4} \right]
\end{equation}
and diffusion coefficient for molar fractions
\begin{equation}
	D^{*}_i = \frac{1}{X_i} \frac{1-Y_i}{\sum_{j\neq i} X_j/ \mathcal{D}_{ij}}
	= \frac{W_i}{Y_i W} \frac{1-Y_i}{\sum_{j\neq i} X_j/ \mathcal{D}_{ij}},
	\label{eq:ma_difcoef_raw}
\end{equation}
with $X_i=c_i/\sum_j c_j$ and $Y_i = W_i c_i / \rho$ denoting molar and mass fractions respectively. $\mathcal{D}_{ij}$ denotes binary diffusion coefficient between species $i$ and $j$.

Molar enthalpies $h_i$ are computed using the thermodynamic data, coded in NASA polynomial format \cite{svehla1962rep}.

Please note that Eqs.~(\ref{eq:cons_species_1d}),(\ref{eq:dens}) imply the conventional continuity equation,
\begin{equation}
	\frac{\partial}{\partial t}\left(r^2 \, \rho \right)
	+ \frac{\partial}{\partial r} \left( r^2 u \rho \right)
	= 0.
\label{eq:cons_mass_1d}
\end{equation}


Our previous studies were concerned with the existence of a critical folding factor, when the deflagration flame propagation cannot be sustained \cite{koksharov18deflagration, kagan2017parametric}, and the role of detailed chemical reaction under such conditions \cite{bykov18ddt}, which assumes an application of a constant flame  folding ratio. However, according to seminal Gostintsev’s experimental observation \cite{gostintsev1988} the flame folding in spherical geometry depends strongly on the flame radius:
\begin{equation}
	\Sigma = \frac{3 A^{2/3} R_f^{1/3}}{2 u_p},
	\label{eq:sigma_radius}
\end{equation}
where $u_p$ is planar flame speed in relation to burned gas for $\Sigma=1$. The factor $A$ depends on mixture composition and pressure. For reaction \ce{2 H2 + O2} under pressure of 1\,bar: $A=2500\,\text{m}^{1/3}/\text{s}^{1/2}$ \cite{gostintsev1988} and $u_p = 90.5\,\text{m}/\text{s}$ \cite{koksharov18deflagration}.

The Maas-Warnatz mechanism \cite{maas88} containing 8 species and 38 elementary reactions is used.

\section{Initial and boundary conditions}
In this setup, a sufficiently large computational domain is chosen, so that the acoustic waves do not disturb the solution by the reflection from the opposite boundary. For the current case, the domain $ 35.1\,\text{mm} \leq r \leq 100\,\text{m}$ (see Eq.~(\ref{eq:bnd_cond})) was chosen. This domain allows a simulation time of about $190$\,ms until acoustic disturbances reach the opposite side, taking into account the speed of sound in the fresh mixture of $540$\,m/s. The position of the left boundary $R_0 = 35.1\,\text{mm}$ is chosen so that the folding factor starts with $\Sigma=1$, according to Eq.\,(\ref{eq:sigma_radius}).
\begin{equation}
\left\{
	\begin{array}{l}
		u(t,R_0) = u(t,R_{\infty}) = 0, \\
		\dfrac{\partial}{\partial r} T(t,R_0) = \dfrac{\partial}{\partial r} T(t,R_{\infty}) = 0, \\
		\dfrac{\partial}{\partial r} c_i(t,R_0) = \dfrac{\partial}{\partial r} c_i(t,R_{\infty}) = 0, ~~\forall~ i=1,..,n_s,
	\end{array}
\right.	
\label{eq:bnd_cond}
\end{equation}

where $R_0 = 35.1\,\text{mm}$ and $R_{\infty} = 100\,\text{m}$.

The ignition is initiated with a hot spot, keeping the initial pressure constant in space.
\begin{equation}
	\left\{
	\begin{array}{l}
		u_0 = 0, \\
		T_0(r) = 300~\mathrm{K} + (1700~\mathrm{K}) e^{-\left( \left(r-R_0\right) / l) \right)^2 }, \\
		c_{0,i}(r) = X_{i,0} \frac{1~\mathrm{bar}}{R^0 T_0(r)}, ~~~~ i = 1,...\;,n_s,\\
	\end{array}
	\right.		
\end{equation}
where hot spot width is $l=0.12$\,mm, which provides a minimal energy deposition.

\section{Numerical method}
The solution is integrated with the Adaptive Pseudo Spectral Element Method (APSEM) suggested in \cite{koksharov18deflagration, bykov2018cfd}. The main idea of the method is to apply the pseudo-spectral method within a particular computational cell with a dynamically adaptive mesh generation via a priori user specified error control and by employing matching procedure \cite{boyd89}. The computational domain is split up into a number of sub-domains while matching the solution $\mathbf{u}$ and its first derivative, e.g. $\partial(\mathbf{u})/\partial x$, at the boundaries, i.e. $\mathbf{u} = \left( u, T, c_1, ..., c_{n_s} \right)$.

The numerical method can be summarised as follows.
The solution domain $r \in [R_0, R_{\infty}]$ on the time interval $t \in [t_a, t_b]$, where $t_0 \leq t_a < t_b \leq t_{\infty}$, is split up into a number of sub-domains
\begin{equation}
	R_0 < \dotsc < r_l < \dotsc < R_{\infty}, ~ l = 1, 2,..\,, (N_r - 1),
\end{equation}
and within each domain $l$, $r \in [r_l, r_{l+1}]$, the solution is approximated with a weighted sum of $M_l$ basis functions
\begin{equation}
	{\mathbf{u}}^l(t,r) = \sum_{i=0}^{M_l-1} \hat{\mathbf{u}}^l_i \, \varphi_i(t,r).
	\label{eq:apsm:fapprox}
\end{equation}
where $\hat{\mathbf{u}}_i$ are corresponding time dependent spectral coefficients.
The number of subdomains $N_r$ and the approximation order $M_l$ are determined simultaneously with the solution $\mathbf{u}$ within iterations for a one time step. The size of the time step and spacial adaptation depend on user specified tolerances (ATOL, RTOL), so that the maximal error does not exceed the specified values.

\subsection{Solution within single numerical domain}
Firstly, a solution with a global numerical domain is presented. Although the main component of the proposed method is the pseudo spectral method \cite{boyd89}, the developed framework allows to easily adapt other spectral methods like the Galerkin weighted residual method. In order to derive the numerical method, the conservation equations can be written in the following form, with corresponding boundary and initial conditions:
\begin{equation}
	\left\{
	\begin{array}{lll}
		\frac{\partial}{\partial t}\left[ \mathbf{f}_t(t, r, \mathbf{u}) \right]
		+ \frac{\partial}{\partial r}\left[ \mathbf{f}_r(t, r, \mathbf{u}, \mathbf{u}') \right]
		= \mathbf{q}(t, r, \mathbf{u}, \mathbf{u}'), 
			& t \in \left[ t_0, \, t_{\infty} \right],
			& r \in \left[ R_0, \, R_{\infty} \right] \\
									
		\frac{\partial}{\partial t}\left[ \mathbf{f}_0(t, \mathbf{u}) \right] = \mathbf{q}_0(t, r, \mathbf{u}, \mathbf{u}'),
			& t \in \left[ t_0, \, t_{\infty} \right]
			& r = R_0 \\
			
		\frac{\partial}{\partial t}\left[ \mathbf{f}_{\infty}(t, \mathbf{u}) \right] = \mathbf{q}_{\infty}(t, r, \mathbf{u}, \mathbf{u}'),
			& t \in \left[ t_0, \, t_{\infty} \right],
			& r = R_{\infty} \\

		\mathbf{u}(t_0, r) = \mathbf{u}_0(r),
			& r \in \left[ R_0, \, R_{\infty} \right] \\
			
	\end{array}
	\right. ,
	\label{method:eq:cons}
\end{equation}

where $\mathbf{f}_t, \, \mathbf{f}_x, \, \mathbf{q}, \, \mathbf{f}_0, \, \mathbf{q}_0, \, \mathbf{f}_{\infty}, \, \mathbf{q}_{\infty}$
 are given vector functions and $\mathbf{u}$
  is the solution vector, $\mathbf{u}'=\frac{\partial}{\partial r} \mathbf{u}$, i.e.
  \begin{equation}
  	\mathbf{f}_t=r^2 \left(
  	\begin{matrix}
  	\rho u \\ \rho e \\ c_i   		
  	\end{matrix}
	\right),
  \end{equation} 
  \begin{equation}
  	\mathbf{f}_r=r^2 \left(
	  	\begin{matrix}
  		\rho u^2 + \tau_{r r} \\
  		u \left( \rho e + P + \tau_{r r} \right) + \sum_{i=1}^{n_s} h_i {j}^d_{i} - \lambda \dfrac{\partial T}{\partial r} \\
  		u c_i + {j}^d_{i}
  	\end{matrix}
	\right),
  \end{equation} 
  \begin{equation}
  	\mathbf{q}=r^2 \left(
  	\begin{matrix}
  	-\dfrac{\partial P}{\partial r} + \dfrac{1}{r} \left( \tau_{\theta \theta} + \tau_{\varphi \varphi}\right) \\
	 0 \\
	 \Sigma^2 \dot{\omega}_i   		
  	\end{matrix}
	\right),	
  \end{equation} 
  \begin{equation}
  	\mathbf{f}_0=\mathbf{f}_{\infty}=\mathbf{0},	
  \end{equation} 
  \begin{equation}
  	\mathbf{q}_0=\mathbf{q}_{\infty}= \left(
  	\begin{matrix}
  	 u \\
	 \dfrac{\partial}{\partial r}T \\
	 \dfrac{\partial}{\partial r}c_i
  	\end{matrix}
	\right).
  \end{equation} 
  
The approximation of the solution $\mathbf{u}$ on an interval $t \in [t_a, t_b]$ and $r \in [r_a, r_b]$ is being sought as a weighted sum of $M_l$ basis functions, see Eq.~(\ref{eq:apsm:fapprox}). It is also required that $\mathbf{u}$ is Lipschitz continuous on the interval $t \in [t_a, t_b]$ and $r \in [r_a, r_b]$. It is also required that vector functions $\mathbf{f}_t, \, \mathbf{f}_r, \, \dot{\mathbf{q}}$ are approximated with the same set of basis functions:
\begin{equation}
	\mathbf{f}(t, r) = \sum_{i=0}^{M_l-1} \hat{\mathbf{f}}_i \, \varphi_i(t, r).
	\label{method:eq:fn_approx}
\end{equation}
Note, for an arbitrary large $M_l$ the approximation leads to the function itself \cite{boyd89}. 

Using a set of given collocation points $(t_i, r_i)$, the relation of $M_l=M_l^t M_l^r$ spectral coefficients and the solution approximation on $M_l^p=(M_l^t-1) M_l^r$ collocation points reads as follows:
\begin{equation}
	\left\{
	\begin{array}{l}
		\sum\limits_{j=0}^{M_l-1} \hat{\mathbf{f}}_{t,j} \, \dot{\varphi}_j(t_0, r_i) = \mathbf{q} \big( t_0, r_i, \mathbf{u}_{0}(r_i),  \mathbf{u}'_{0}(r_i)\big) - \dfrac{\partial}{\partial r} \mathbf{f}_x \big( t_0, r_i, \mathbf{u}_{0}(r_i),  \mathbf{u}'_{0}(r_i)\big),\\
		~~~~~~~~~~~i= 0,...,M_l^r-1, \\
		\sum\limits_{j=0}^{M_l-1} \hat{\mathbf{f}}_{t,j} \, \varphi_j(t_i, r_i) = \mathbf{f}_t \big( t_i, r_i, \mathbf{u}(t_i, r_i),  \mathbf{u}'(t_i, r_i)\big), \\
		~~~~~~~~~~~i=k_t M_l^r + k_r,~\forall ~ k_t = 0,...,M_l^t-2 \, \wedge \, k_r = 0,...,M_l^r-1, \\
	\end{array}
	\right.
	\label{method:eq:fn_fit_ft}
\end{equation}
and
\begin{equation}
	\left\{
	\begin{array}{l}
		\sum\limits_{j=0}^{M_l-1} \hat{\mathbf{f}}_{x,j} \, \varphi'_j(t_0, r_i) = \dfrac{\partial}{\partial r} \mathbf{f}_x \big( t_0, r_i, \mathbf{u}_{0}(r_i),  \mathbf{u}'_{0}(r_i)\big),
		~~i= 0,...,M_l^r-1. \\
		\sum\limits_{j=0}^{M_l-1} \hat{\mathbf{f}}_{x,j} \, \varphi_j(t_i, r_i) = \mathbf{f}_x \big( t_i, r_i, \mathbf{u}(t_i, r_i),  \mathbf{u}'(t_i, r_i)\big), \\
		~~~~~~~~~~~i=k_t M_l^r + k_r,~\forall ~ k_t = 0,...,M_l^t-2 \, \wedge \, k_r = 0,...,M_l^r-1, \\
	\end{array}
	\right.
	\label{method:eq:fn_fit_fx}
\end{equation}

Eqs.~(\ref{method:eq:fn_approx}), (\ref{method:eq:fn_fit_ft}) and (\ref{method:eq:fn_fit_fx}) can be more conveniently written in matrix-vector form:
\begin{equation}
	\widetilde{\mathbf{f}}_{t} = \hat{\mathbf{f}}_{t} \Phi,
\end{equation}
\begin{equation}
	\hat{\mathbf{f}}_{t} \mathrm{P}_t = \left(\widetilde{\mathbf{q}}^{0} - \tfrac{\partial}{\partial r}\widetilde{\mathbf{f}}^0_x,~ \widetilde{\mathbf{f}}_{t} \right),
\end{equation}
and
\begin{equation}
	\hat{\mathbf{f}}_{x} \mathrm{P}_x = \left( \tfrac{\partial}{\partial r}\widetilde{\mathbf{f}}^0_x,~ \widetilde{\mathbf{f}}_{x}\right),
\end{equation}
where $\widetilde{\mathbf{f}}_{t}$, $\widetilde{\mathbf{f}}_{x}$, $\widetilde{\mathbf{q}}$ are the solution approximations on the specified $M_l^p$ collocation points and $\widetilde{\mathbf{f}}^0_{t}$, $\widetilde{\mathbf{f}}^0_{x}$, $\widetilde{\mathbf{q}}^0$ are initial values at collocation points for $t=t_a$, such that
\begin{equation}
	\widetilde{\mathbf{f}}_{t,i} = 
	\mathbf{f}_t \big( t_i, r_i, \mathbf{u}(t_i, r_i),  \mathbf{u}'(t_i, r_i)\big),
\end{equation}
\begin{equation}
	\widetilde{\mathbf{f}}_{x,i} = 
	\mathbf{f}_x \big( t_i, r_i, \mathbf{u}(t_i, r_i),  \mathbf{u}'(t_i, r_i)\big),
\end{equation}
\begin{equation}
	\widetilde{\mathbf{q}}_{i} = 
	\mathbf{q} \big( t_i, r_i, \mathbf{u}(t_i, r_i),  \mathbf{u}'(t_i, r_i)\big),
\end{equation}
\begin{equation}
	\widetilde{\mathbf{f}}^0_{t,i} = 
	\mathbf{f}_t \big( t_i, r_i, \mathbf{u}_0(r_i),  \mathbf{u}_0'(r_i)\big),
\end{equation}
\begin{equation}
	\widetilde{\mathbf{f}}^0_{x,i} = 
	\mathbf{f}_x \big( t_i, r_i, \mathbf{u}_0(r_i),  \mathbf{u}_0'(r_i)\big),
\end{equation}
\begin{equation}
	\widetilde{\mathbf{q}}^0_{i} = 
	\mathbf{q}_0 \big( t_i, r_i, \mathbf{u}_0(r_i),  \mathbf{u}_0'(r_i)\big),
\end{equation}

Then time and space derivatives can be computed from the collocation points as
\begin{equation}
	\dot{\widetilde{\mathbf{f}}}_t 
	= \left(\widetilde{\mathbf{q}}_{0} - \tfrac{\partial}{\partial r}\widetilde{\mathbf{f}}^0_x,~ \widetilde{\mathbf{f}}_{t} \right) P_t^{-1} \dot{\Phi} 
	= \left(\widetilde{\mathbf{q}}_{0} - \tfrac{\partial}{\partial r}\widetilde{\mathbf{f}}^0_x,~ \widetilde{\mathbf{f}}_{t} \right)  \dot{\mathrm{H}},
\end{equation}
and
\begin{equation}
	{\widetilde{\mathbf{f}}}'_x 
	= \left( \tfrac{\partial}{\partial r}\widetilde{\mathbf{f}}^0_x,~ \widetilde{\mathbf{f}}_{x}\right) P_x^{-1} \Phi' 
	= \left( \tfrac{\partial}{\partial r}\widetilde{\mathbf{f}}^0_x,~ \widetilde{\mathbf{f}}_{x}\right) \mathrm{H}'.
\end{equation}

Similarly, the first derivative of a one dimensional time function $\mathbf{f}_t=\sum_i \hat{\mathbf{f}}_t \varphi_i (t)$ is $\dot{\widetilde{\mathbf{f}}}_t=\left( \tfrac{\mathrm{d}}{\mathrm{d} t}\widetilde{\mathbf{f}}^0_t,~ \widetilde{\mathbf{f}}_{t}\right) \mathrm{G}'_t$.

Using a set of test functions $\psi_i$ a weak formulation can be written
\begin{equation}
	\int_{t_a}^{t_b} \int_{r_a}^{r_b} \left[
	\frac{\partial}{\partial t}\left( \mathbf{f}_t \right) + \frac{\partial}{\partial r}\left( \mathbf{f}_r \right)
	\right] \, \psi_i \, \mathrm{d}t \, \mathrm{d}r
	= \int_{t_a}^{t_b} \int_{r_a}^{r_b} \mathbf{q} \, \psi_i \, \mathrm{d}t \, \mathrm{d}r, \: \forall \: i=0,..\,, M_l^p-1.
	\label{method:eq:weak}
\end{equation}

Introducing eq.~(\ref{method:eq:fn_approx}) in eq.~(\ref{method:eq:weak}) and choosing test functions from the space of Dirac-delta functions $\psi_i(t,r) = \delta(t-t_i)\delta(r-r_i)$ one obtains a discrete form of conservation equations:
\begin{equation}
\left\{
\begin{array}{l}	
	\sum\limits_{j=M_l^r}^{M_l^p-1} \left(  
		\widetilde{\mathbf{q}}_{j} 
		- \widetilde{\mathbf{f}}_{t,j} \, \dot{h}_{j+M_l^r,i} 
		- \widetilde{\mathbf{f}}_{x,j} \, {h}'_{j+M_l^r,i}
		\right) \\
	\begin{array}{r}
		~~~= \sum\limits_{j=0}^{M_l^r-1} \Bigg( 
		\left(\widetilde{\mathbf{q}}_{j}^0 - \tfrac{\partial}{\partial r}\widetilde{\mathbf{f}}_{x,j}^0\right) \dot{h}_{j,i}
		+ \widetilde{\mathbf{f}}_{t,j}^0 \dot{h}_{j+M_l^r,i}
		+ \tfrac{\partial}{\partial r} \widetilde{\mathbf{f}}_{x,j}^0 {h}'_{j,i}
		+ \widetilde{\mathbf{f}}_{x,j}^0 {h}'_{j+M_l^r,i}
		\Bigg), \\
	 i=k_t M_l^r + k_r, \: \forall \: k_t=1,...\,, M_l^t-2\, \wedge \, k_r=1,...\,,M_l^r-1, 
	\end{array}
	\\
	\sum\limits_{j=1}^{M_l^t-2} \left(  
		\widetilde{\mathbf{q}}_{0,j} 
		- \widetilde{\mathbf{f}}_{0,j} \, {g}^t_{j+1,i}
		\right) 
		=  
		\widetilde{\mathbf{q}}_{0,0}^0 {g}^t_{0,0}
		+ \widetilde{\mathbf{f}}_{0,0}^0 {g}^t_{1,0},~~\forall ~i=1,...\,,M_l^t-2,	
	\\
	\sum\limits_{j=1}^{M_l^t-2} \left(  
		\widetilde{\mathbf{q}}_{\infty,j} 
		- \widetilde{\mathbf{f}}_{\infty,j} \, {g}^t_{j+1,i}
		\right) 
		=  
		\widetilde{\mathbf{q}}_{\infty,0}^0 {g}^t_{0,0}
		+ \widetilde{\mathbf{f}}_{\infty,0}^0 {g}^t_{1,0},~~\forall ~i=1,...\,,M_l^t-2.
\end{array}	
\right.
\label{mathod:eq:discrete}
\end{equation}

If the number of basis function in time direction is $M_l^t=3$, the time marching scheme is equivalent to the well known trapezoidal (Crank-Nicolson) method (see e.g. \cite{Quarteroni00}), which is $O(2)$ accurate. In the current study we use: for space -- Chebyshev polynomials of the first kind of variable order ($M_l^r \leq 16$) with Gauss-Lobatto collocation points \cite{boyd89}, and for time -- power series of the second order ($M_l^t=3$) with equidistant points, i.e. for time interval $t \in [t_a, t_b]$ and space -- $r \in [r_l, r_{l+1}]$ it reads as
\begin{equation}
	\varphi_i(t, r) = \varphi_j^t(t) \varphi_k^r(r) = \left( \frac{t-t_a}{t_b - t_a} \right)^j \, T_k\left( 2\frac{r-r_l}{r_{l+1} - r_l} - 1 \right)	,
\end{equation}
where $i=j M^r_l + k$, $T_k$ is $k$-th Chebyshev polynomial of the first kind and $M_l = M^t_l M^r_l$ with $M^t_l$ is the number of basis functions in time direction and $M^r_l$ is the number of basis functions in space direction. Note, however, the presented algorithm can be used with other types of basis functions. For an elaborated description of the method the reader is referred to \cite{bykov2018cfd}.

The system (\ref{mathod:eq:discrete}) is a non-linear algebraic equation, which is solved implicitly with Newton's method (see e.g. \cite{Quarteroni00}).

\subsection{Domain matching and boundary conditions}


The solution in time is being sought one time-interval after the other:  the initial condition $\mathbf{u}(t_k)$  on the interval $t \in [t_k, t_{k+1}] $ is provided from the solution of the previous interval $t \in [t_{k-1}, t_{k}] $.

The subdomains in space are matched with the following boundary conditions:
\begin{equation}
	\left\{
	\begin{array}{l}
		\mathbf{u}_l(t,r_{l-1}) = \mathbf{u}_{l-1}(t,r_{l-1}) \\	
		\mathbf{u}_l(t,r_{l}) = \mathbf{u}_{l+1}(t,r_{l}) \\	
		\frac{\partial}{\partial x}\mathbf{u}_l(t,r_{l-1}) = \frac{\partial}{\partial r}\mathbf{u}_{l-1}(t,r_{l-1}) \\		
		\frac{\partial}{\partial x}\mathbf{u}_l(t,r_{l}) = \frac{\partial}{\partial r}\mathbf{u}_{l+1}(t,r_{l}), \\
	\end{array}
	\right. ,
	\label{method:eq:bnd_match}
\end{equation}
where $\mathbf{u}_l$ is solution approximation within domain $l$. Note, first and last collocation points (Gauss-Lobatto) are placed exactly at the boundary. The evolution of these points (\textit{knots}) is governed by eqs.~(\ref{method:eq:bnd_match}), whilst points within a space interval (\textit{cell}) are governed by PDE eqs.~(\ref{method:eq:cons}).

\subsection{Grid adaptation}
In order to introduce the grid adaptation strategy, the error control strategy is first presented. The absolute interpolation error of the state variable $u_i$ for the number of basis functions $M$ is limited by the factor of 2 of the series last coefficient \cite{boyd89}, but even in converging series, under particular conditions, some coefficients can be naturally zeros, whilst others are non-zero. For these reasons the following approximation of the absolute error of variable $u_i$ has been suggested and used:
\begin{equation}
	e^{\mathrm{abs}}_i = \sqrt{ \hat{u}^2_{i,M} + \hat{u}^2_{i,M-1}}, ~~\forall~ i=0,...,n_v-1,
\end{equation}
where $\hat{u}_{i,k}$ is the amplitude of the $k$-th coefficient of the series expansion. The last equation is used in order to deal with spurious zeros of spectral coefficients. Then the normalized relative error is defined as:
\begin{equation}
	E = \displaystyle{\max_{i=1,...,n_v}} 
	\frac
	{e^{\mathrm{abs}}_i}
	{ \mathrm{RTOL}_i \big( \mathrm{ATOL}_i + \left|\hat{u}_{i,0}\right| \big) }.
\label{eq:apsm:error}
\end{equation}

When eq.~(\ref{eq:apsm:error}) is applied for each time slice, corresponding to time collocation points, the measure of space error $E^r$ is derived. When the equation is applied for each space slice, corresponding to space collocation points, the measure of time error $E^t$ is computed.

The time adaptation is reflected in the size of time interval, while keeping the order of polynomial along time direction constant, e.g. $M_l^t=3$. The procedure of the time step adaptation is covered by standard numerical recipes (see e.g. \cite{Quarteroni00}).

The adaptation strategy in space direction $r$ consists of two parts: one is to adjust the polynomial order within one computational cell; the other is to split/join cells. The first strategy becomes rapidly computationally expensive for large polynomial orders, e.g. for $M_l^r>16$, due to the computationally expensive inversion of large dense matrices. However, the error of approximation is sinking exponentially with the polynomial order. The cell splitting creates more collocation points and is hampered by memory reallocation, compared to the polynomial order increase, but it is outweighed by a faster global solution of the linear system and allows to avoid spectral blocking when the relatively stiff behaviour of the solution takes place. 

If the accuracy within a cell $l$ is insufficient, i.e. $E_l^r > 1$, the adaptation consists of two stages:
\begin{enumerate}
	\item 
	Increase polynomial order within cell till it reaches $M_l^r=16$, if the error is still above unity;
	\item
	Split computational cell.
\end{enumerate}
If the accuracy within the cell $l$ exceeds requirements (specified tolerances), i.e. $E^r_l < 0.01$, the adaptation consist again of two stages:
\begin{enumerate}
	\item 
	Decrease polynomial order within the cell until it reaches $M_l^r=4$, if the the error is still below unity;
	\item
	If the adjacent cell has also reached the minimal polynomial order, merge both cells to one cell, while keeping the order $M_l^r=4$.
\end{enumerate}

The specified maximal order $M_l^r=16$, minimal order $M_l^r=4$, minimal normalised error $E_l^r=0.01$ are customisable parameters, which to a large extent depend on a type of polynomial approximation. By the reduction of polynomial order and by joining of computational cells it is important to enforce not only boundary values (imposed automatically) but also gradients at the boundary (see Eq.~\ref{method:eq:bnd_match}).

\section{Results and discussions}
\begin{figure}[htbp]
  \centering
  \includegraphics[width=0.49\columnwidth]{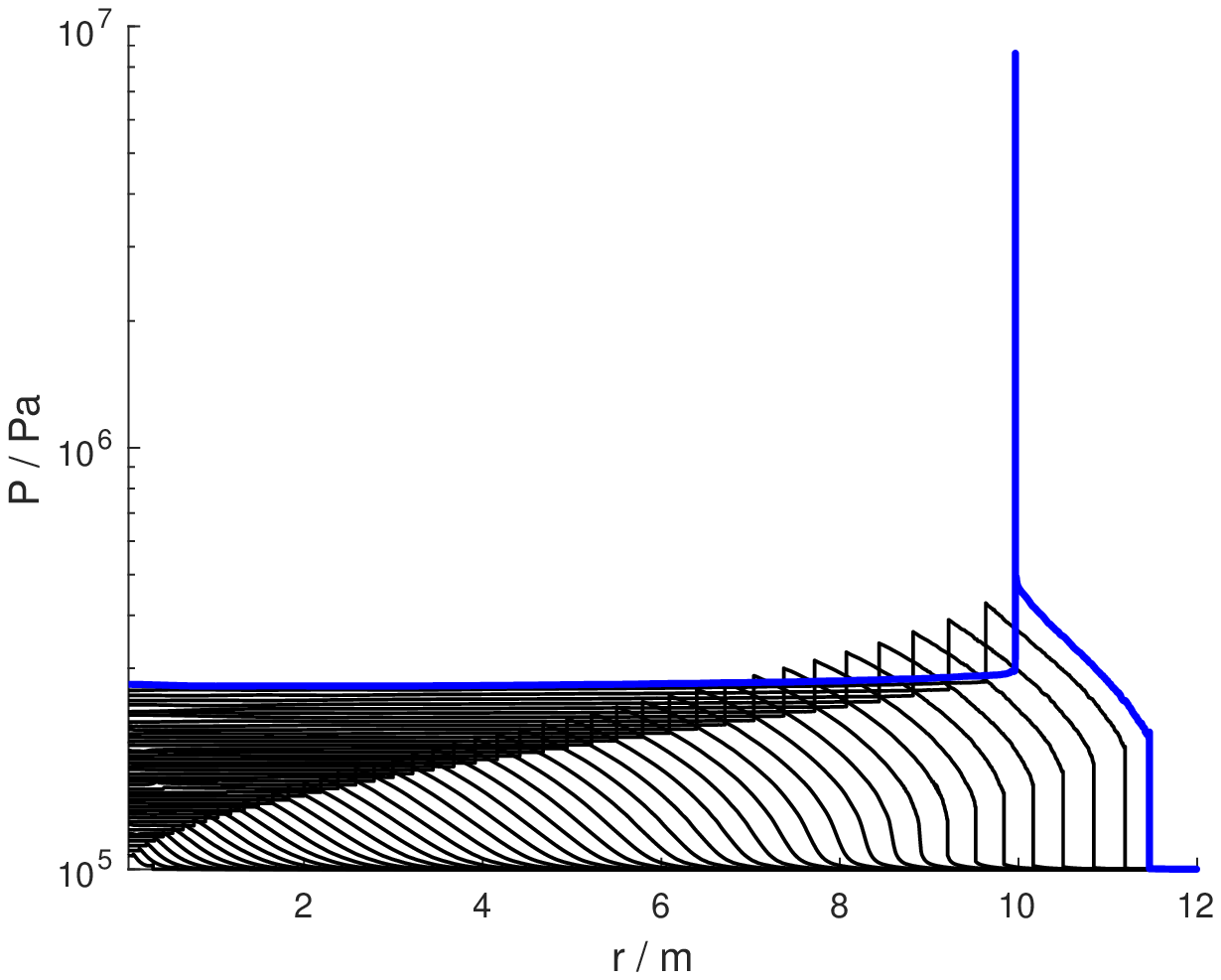}
  \includegraphics[width=0.49\columnwidth]{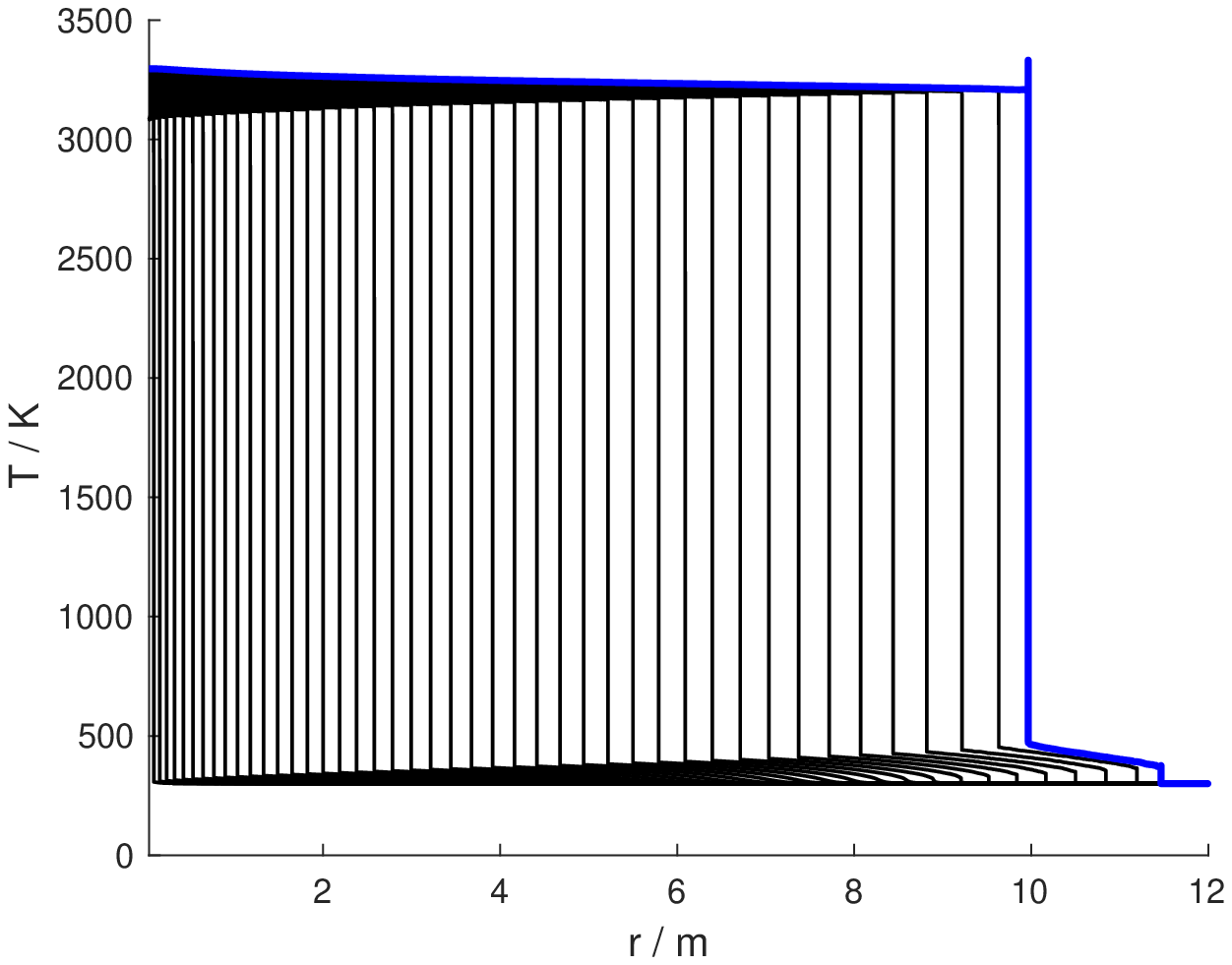}
  \caption{Pressure (left) and temperature (right) profiles every $0.5\,\mu s$. Blue line corresponds to DDT.}
  \label{fig:prof_pt}
\end{figure}

\begin{figure}[htbp]
  \centering
  \includegraphics[width=0.49\columnwidth]{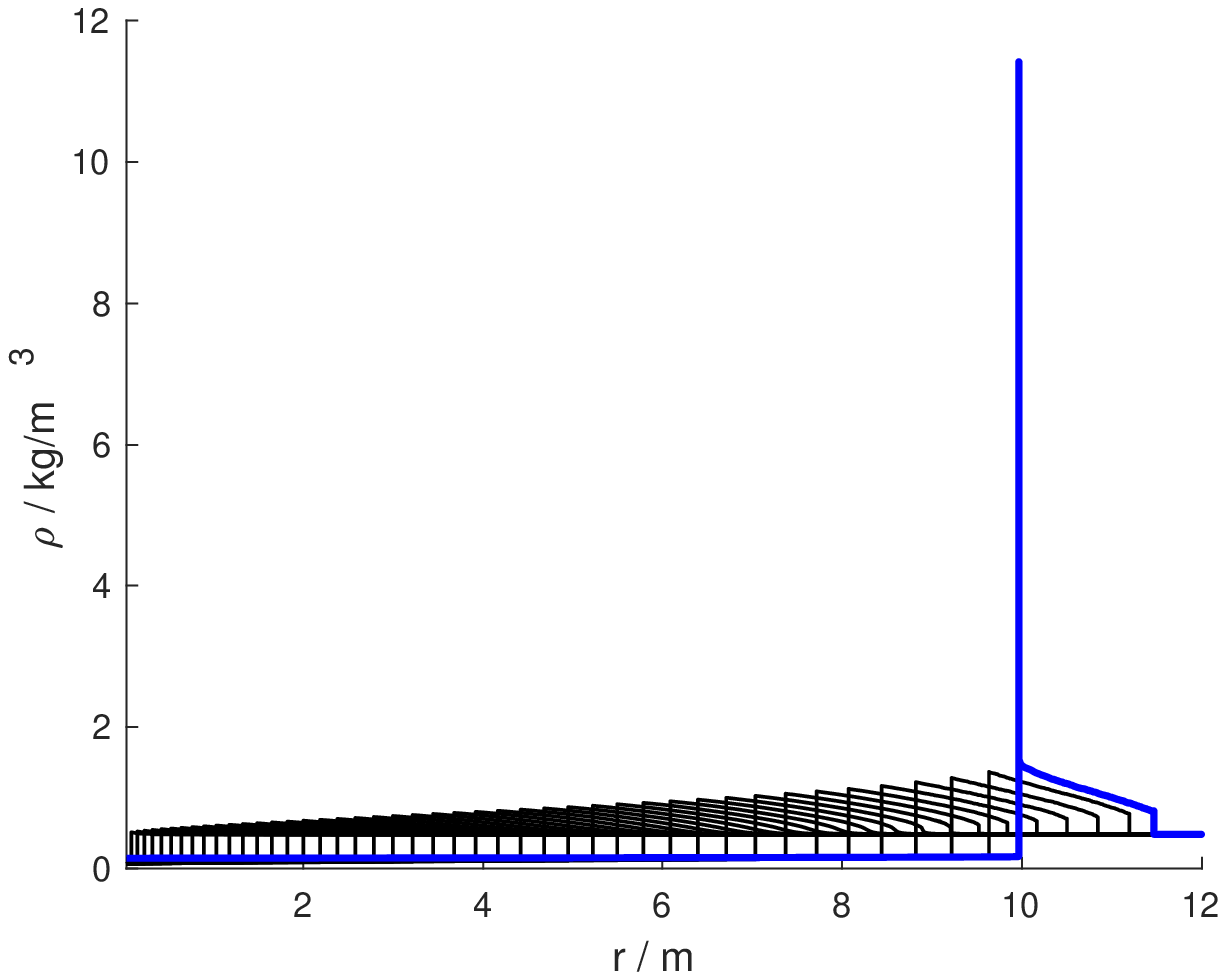}
  \includegraphics[width=0.49\columnwidth]{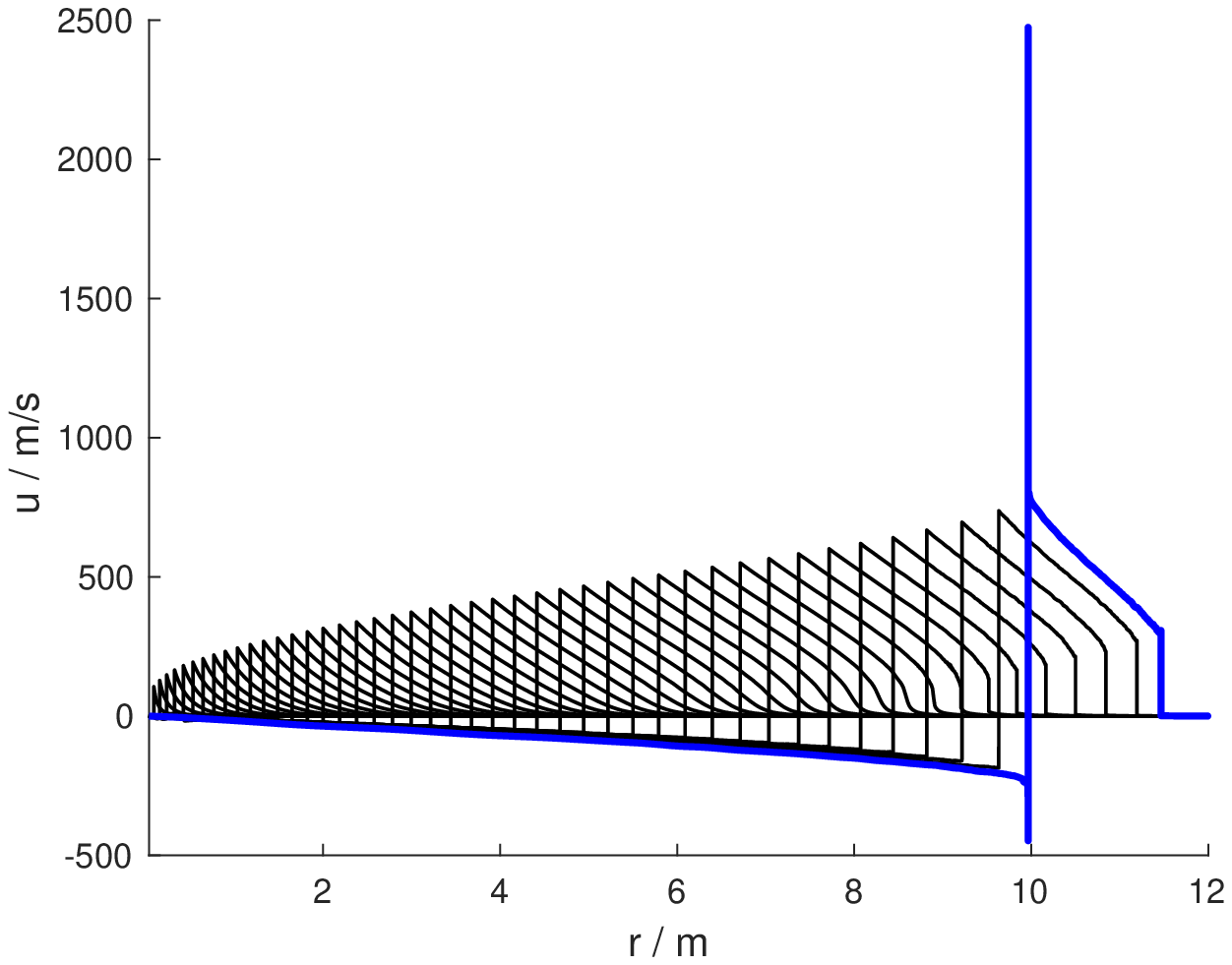}
  \caption{Density (left) and flow velocity (right) profiles every $0.5\,\mu s$. Blue line corresponds to DDT.}
  \label{fig:prof_rhov}
\end{figure}

Fig.~\ref{fig:prof_pt} shows pressure (left) and temperature (right) profiles, plotted every $0.5\,\mu s$, as well as the state at the DDT point. Fig.~\ref{fig:prof_rhov} shows the corresponding density and velocity profiles. The observed transition radius and critical flame folding ratio are $R_{ddt}=9.965$\,m and $\Sigma_{ddt}=6.57$. In our previous work \cite{koksharov18deflagration} we estimated in planar symmetry the transition radius $R^*_{ddt}=8.377$\, and the corresponding critical flame folding ratio $\Sigma^*_{ddt}=6.25$. The latter values as expected are slightly less than in the current spherical case.

\begin{figure}[htbp]
  \centering
  \includegraphics[width=0.49\columnwidth]{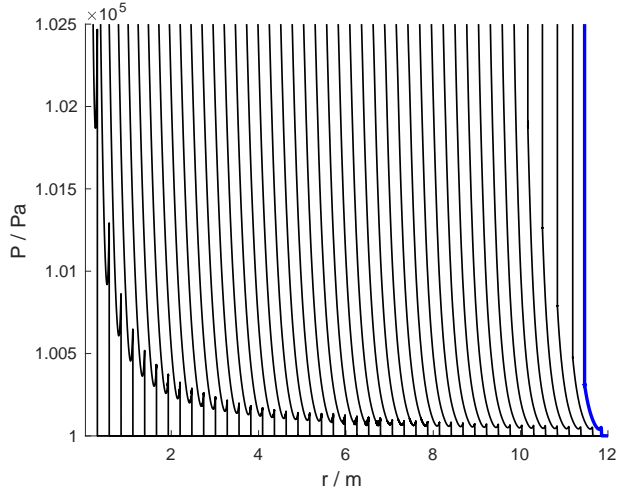}
  \caption{Zoomed pressure profiles for every $0.5\,\mu s$. Blue line corresponds to DDT.}
  \label{fig:prof_zpt}
\end{figure}

Although a smooth ignition was applied (providing ignition energy less by 1\,\% leads to ignition failure), a generated during the ignition shock wave is traveling ahead of the flame. This is shown on the pressure zoomed pressure profile Fig.~\ref{fig:prof_zpt}.

\begin{figure}[htbp]
  \centering
  \includegraphics[width=0.49\columnwidth]{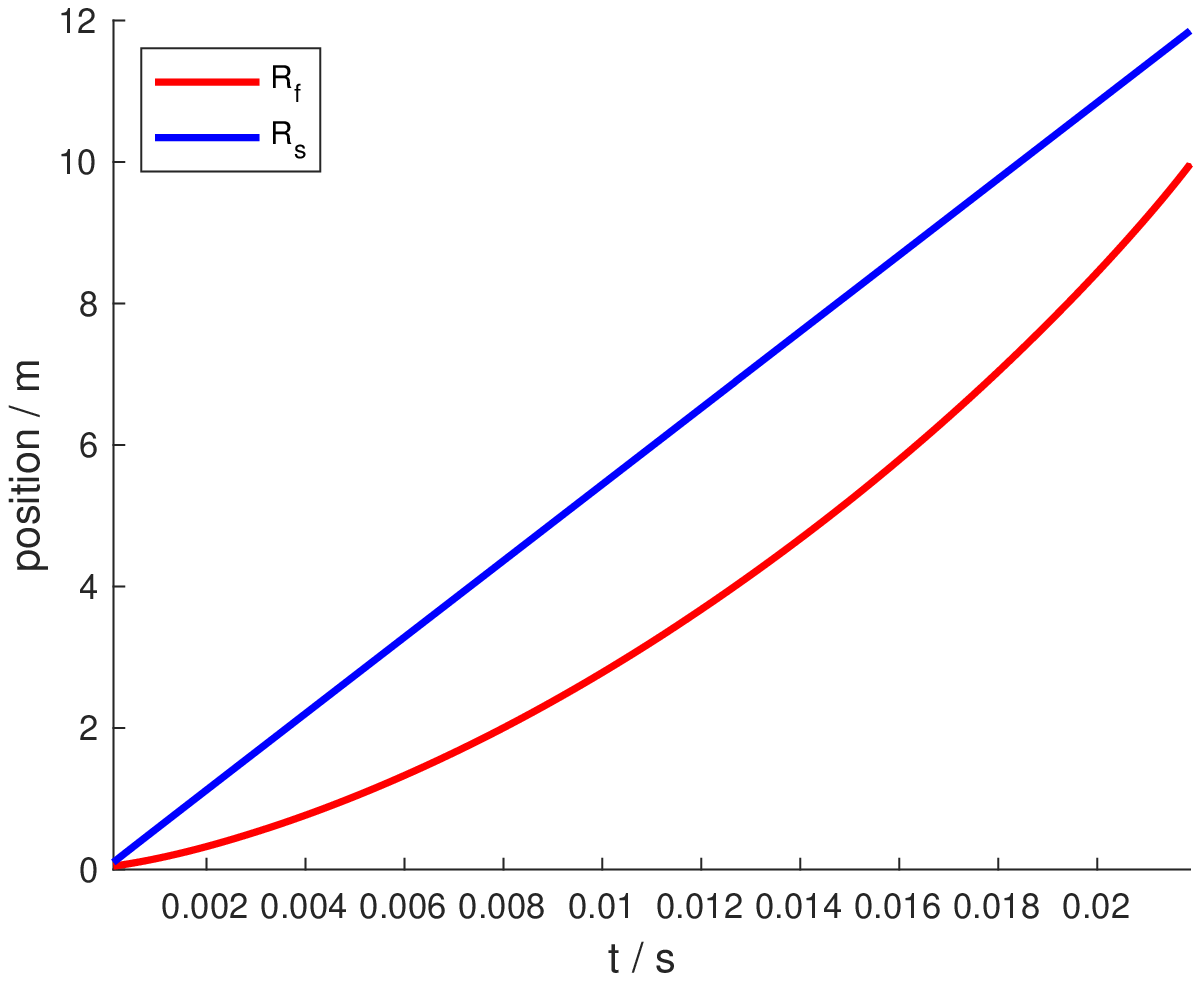}
  \includegraphics[width=0.49\columnwidth]{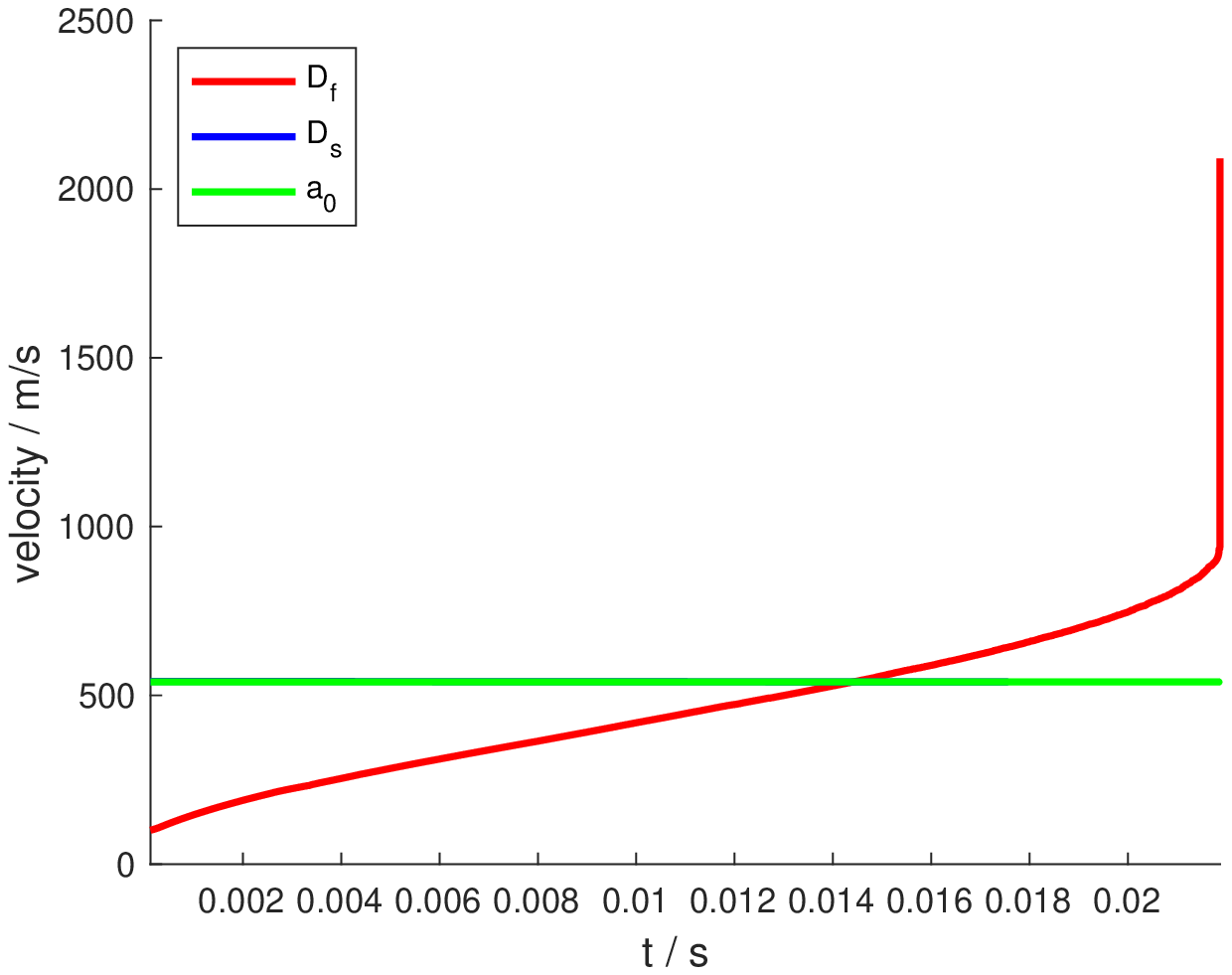}
  \caption{Flame and shock position vs. time (left); flame flame and shock velocity vs. time (right).}
  \label{fig:rf-t}
\end{figure}

Positions of flame front $R_f$ and of the precursor shock $R_s$ at different time instances are shown in Fig. \ref{fig:rf-t} (left). Correspondingly the flame front velocity $D_f$ and the velocity of the precursor shock $D_s$ are shown in Fig. \ref{fig:rf-t} (right). As one may expect, the speed of the precursor shock is close to the speed of sound $a_0$ in the unburned mixture, which is plotted by the green line. Note that these green and blue lines practically coincide. In order to compute the speed of sound the following expression is used \cite{williams88}:
\begin{equation}
	a = \sqrt{\gamma \frac{P}{\rho}},
\end{equation}
where
\begin{equation}
	\gamma = \frac{\sum\limits_i c_i \, {C_p}_i}{\sum\limits_i c_i \left( {C_p}_i - R^0 \right)}
\end{equation}
is the adiabatic index for an ideal gas and ${C_p}_i$ is the molar heat capacity at constant pressure.

\begin{figure}[htbp]
  \centering
  \includegraphics[width=0.49\columnwidth]{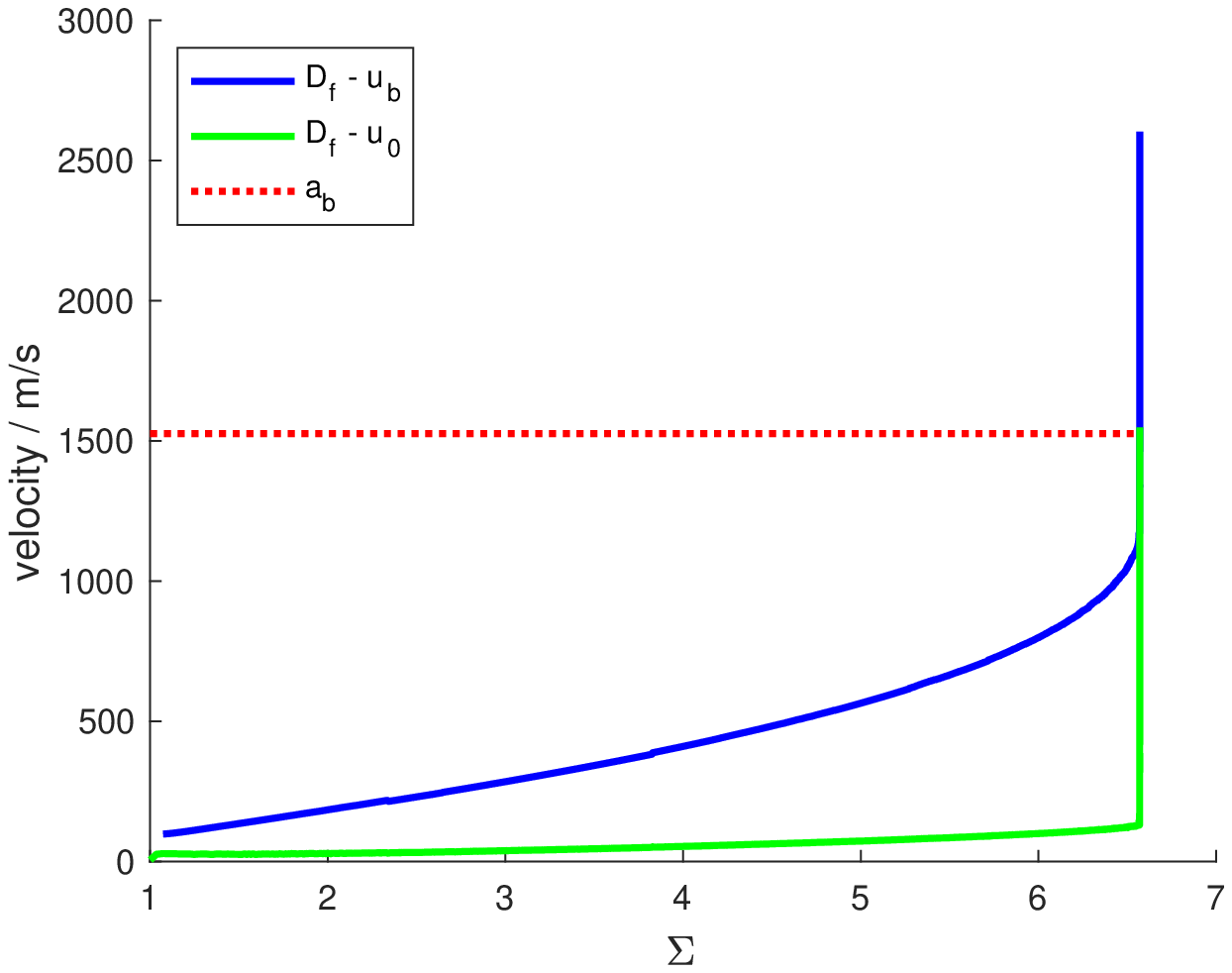}
  \includegraphics[width=0.49\columnwidth]{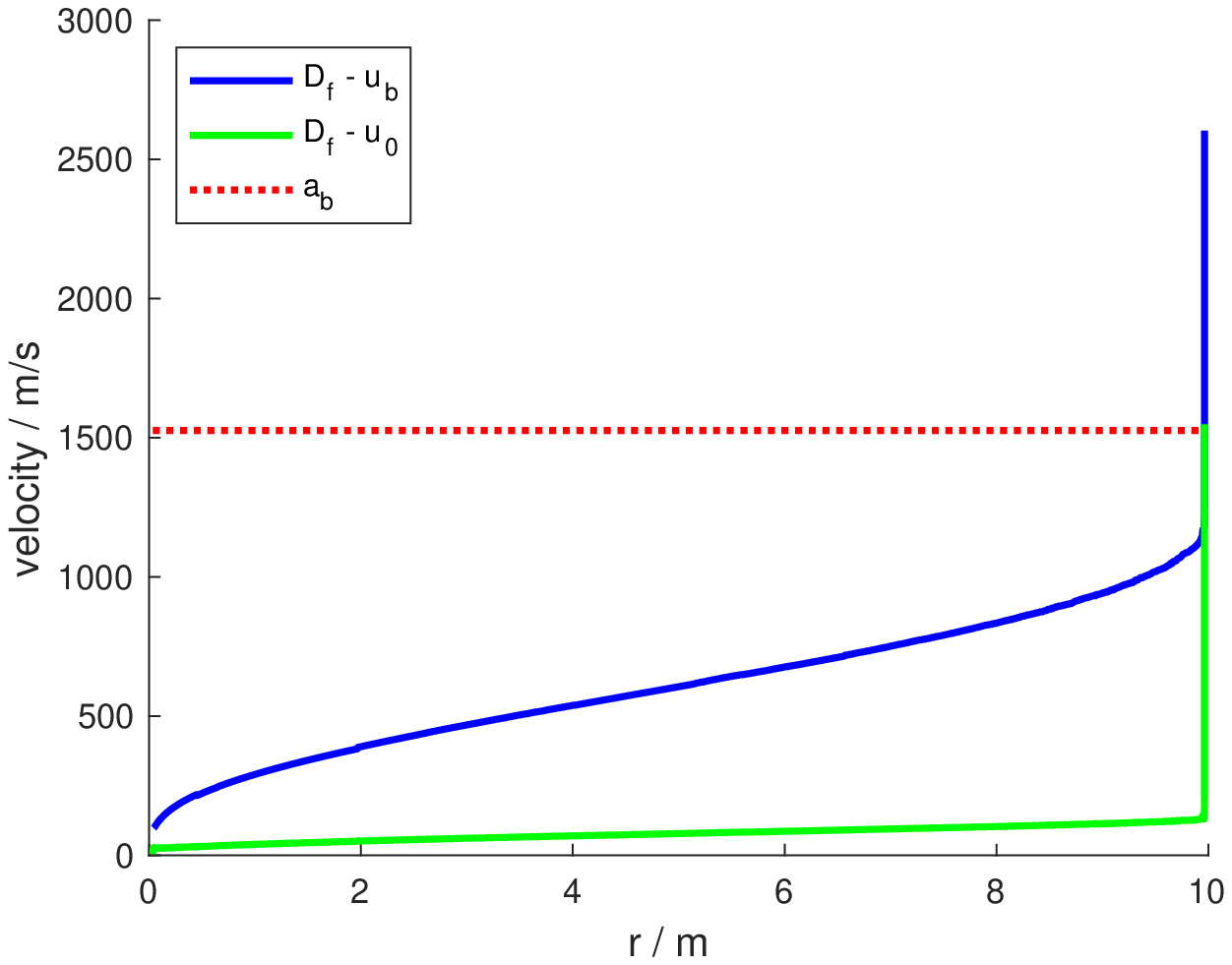}
  \caption{Flame front velocity relative to the entry (green) and to the exit (blue) from the reaction zone, as a function of folding ratio (left) and as a function of flame position (right).}
  \label{fig:dv_r}
\end{figure}

As is readily seen the flame speed $D_f$ increases and may even exceed $a_0$.  Yet, approaching the DDT point the accelerating flame stays behind the precursor shock: $R_{ddt}=10\;\mathrm{m}$ while $R_{s,ddt}=11.5\;\mathrm{m}$.
In Fig.~\ref{fig:dv_r} $u_0=u(R_f + 5\;\mathrm{mm})$, $u_b=u(R_f - 5\;\mathrm{mm})$ corresponding respectively to the approximate entry/exit into/from the reaction zone. Since $D_f - u_b < a_b$, the pre-DDT flame does not reach the threshold of CJ-deflagration. Approaching the transition point $\Sigma$-dependency of $D_f - u_b$ becomes nonlinear. This can be attributed to the elevated pressure and temperature in the compressed region between $R_f$ and $R_s$, conducive to a higher burning velocity.

\clearpage
\section*{Acknowledgement}
These studies were supported by the US-Israel Binational Science Foundation (Grant 2012-057) and the Israel Science Foundation (Grant 335/13).

\thispagestyle{plain}

\bibliographystyle{elsarticle-num}
\bibliography{lit.bib}

\end{document}